\documentclass[unnumsec,webpdf,modern,large,namedate]{oup-authoring-template}% uncomment this line for author year citations and comment the above
\graphicspath{{Fig/}}

\begin{document}

\journaltitle{arXiv}
\DOI{}
\copyrightyear{2026}
\pubyear{2026}
\appnotes{\textbf{Application Note}}

\firstpage{1}

\subtitle{Gene Expression}

\title[EMMA]{EMMA: an R/Bioconductor package to automate tracking of metadata in functional enrichment analyses}

\author[1$\ast$]{Najla Abassi\ORCID{0000-0001-8357-0938}}
\author[1]{Annekathrin Silvia Nedwed \ORCID{0000-0002-2475-4945}}
\author[1,2$\ast$]{Federico Marini\ORCID{0000-0003-3252-7758}}

\authormark{Abassi et al.}

\address[1]{\orgdiv{Institute of Medical Biostatistics, Epidemiology and Informatics (IMBEI)}, \orgname{University Medical Center Mainz}, \state{Mainz}, \country{Germany}}
\address[2]{\orgdiv{Research Center for Immunotherapy (FZI) Mainz},\state{Mainz}, \country{Germany}}

\corresp[$\ast$]{Corresponding authors. \href{mailto:najla.abassi@uni-mainz.de}{najla.abassi@uni-mainz.de} and \href{mailto:marinif@uni-mainz.de}{marinif@uni-mainz.de}}

\received{Date}{0}{Year}
\revised{Date}{0}{Year}
\accepted{Date}{0}{Year}

%\editor{Associate Editor: Name}

%\abstract{
%\textbf{Motivation:} .\\
%\textbf{Results:} .\\
%\textbf{Availability:} .\\
%\textbf{Contact:} \href{name@email.com}{name@email.com}\\
%\textbf{Supplementary information:} Supplementary data are available at \textit{Journal Name}
%online.}

\abstract{
\textbf{Summary:}
Functional enrichment analysis (FEA) is a widely used approach for interpreting high-throughput omics data. However, essential methodological details, such as software versions, analysis parameters, and annotation database releases among others, are often incompletely reported, limiting the reproducibility and transparency of enrichment analyses and complicating the assessment of potentially problematic methodological choices. Here we present EMMA, an R/Bioconductor package that integrates with existing FEA tools and automatically captures provenance metadata, such as annotation metadata, software version, and parameters, during the analysis runtime. Our package provides utilities for accessing and exporting the recorded metadata to facilitate transparent reporting and preserve provenance required for reproducible enrichment analyses. This also enables auditing of the results while remaining compatible with existing Bioconductor workflows. \\
\textbf{Availability and implementation:}
EMMA is available on Bioconductor under the MIT license (\url{https://bioconductor.org/packages/EMMA}), with its development version also available on GitHub (\url{https://github.com/imbeimainz/EMMA}).
}

\maketitle

\section{Introduction}

Functional Enrichment Analysis (FEA) has become a standard downstream step in omics studies, enabling the biological interpretation of candidate features (e.g. genes,  proteins, metabolites) identified through differential analyses \citep{Zhao2023, Xu2024}.

Numerous methods and tools have been developed to perform FEA, with Over-Representation Analysis (ORA) \citep{Khatri2012} and Gene Set Enrichment Analysis (GSEA) \citep{Subramanian2005} being among the most widely used approaches. The diversity of available methods and their implementations means that FEA workflows involve a variety of analytical decisions, including the selection of enrichment methods, gene set database versions, background gene universes, and multiple testing correction methods. These methodological choices can substantially influence the resulting biological conclusions \citep{Timmons2015, Wadi2016}, making transparent and complete reporting of enrichment analyses essential.
However, FEA workflows are often inadequately documented \citep{Wijesooriya2022}. Critical methodological details, such as the software used, the choice of background gene universe or the application of multiple testing correction, are frequently incompletely reported in published studies, compromising reproducibility and interpretation of the results \citep{Wijesooriya2022, chen2023fair, Bora2026}. Recommendations for transparent and reproducible FEA have existed for nearly two decades \citep{YonRhee2008, Hung2011} and continue to be emphasized in more recent best practice guidelines and protocol publications \citep{Chicco2022, Zhao2023, Xu2024, Bora2026}. Nevertheless, methodological inconsistencies remain common in the literature, and manual documentation is particularly difficult because enrichment packages provide heterogeneous interfaces, apply different defaults, and obtain annotation resources through different mechanisms.

In computational science, reproducibility has been established as a fundamental principle for assessing published findings \citep{Peng2011}. In omics research, minimum-information standards can improve the availability and interpretability of results. For example, the MIAME guidelines established structured reporting requirements for microarray experiments and promoted the deposition of data and metadata supporting published analyses \citep{Brazma2001, Brazma2009}.
Automatic metadata capture has also been successfully applied in computational genomics. For instance, omicslog (\url{https://github.com/tidyomics/omicslog}) tracks transformations applied to SummarizedExperiment objects in complex analysis workflows. Similarly, tximeta associates RNA-sequencing quantifications with reference-sequence and annotation provenance \citep{Love2020}. For FEA, Chen et al. proposed minimum metadata requirements of enrichment provenance \citep{chen2023fair}. Despite some tools, such as gprofiler2 \citep{Kolberg2020}, capturing metadata including database versions, these requirements are not always automatically captured across commonly used R enrichment packages.

Here we present EMMA (Enrichment Methods MAtter), an R/Bioconductor package that integrates with existing FEA tools, and automatically captures provenance metadata during analysis runtime. Notably, EMMA preserves the original format of the enrichment results while generating structured, reusable metadata, such as the enrichment method used or input parameters. The resulting provenance record supports transparent reporting, interpretation, and computational reproducibility without requiring any changes to existing workflows, with the only cost being a negligible increase in the size of the result object.

A complete documentation and exemplary application of EMMA is provided in its vignette through Bioconductor (also rendered at \url{https://imbeimainz.github.io/EMMA/}).

\section{Implementation}

\begin{figure*}[!ht]%
\centering
\includegraphics[width=0.95\linewidth]
{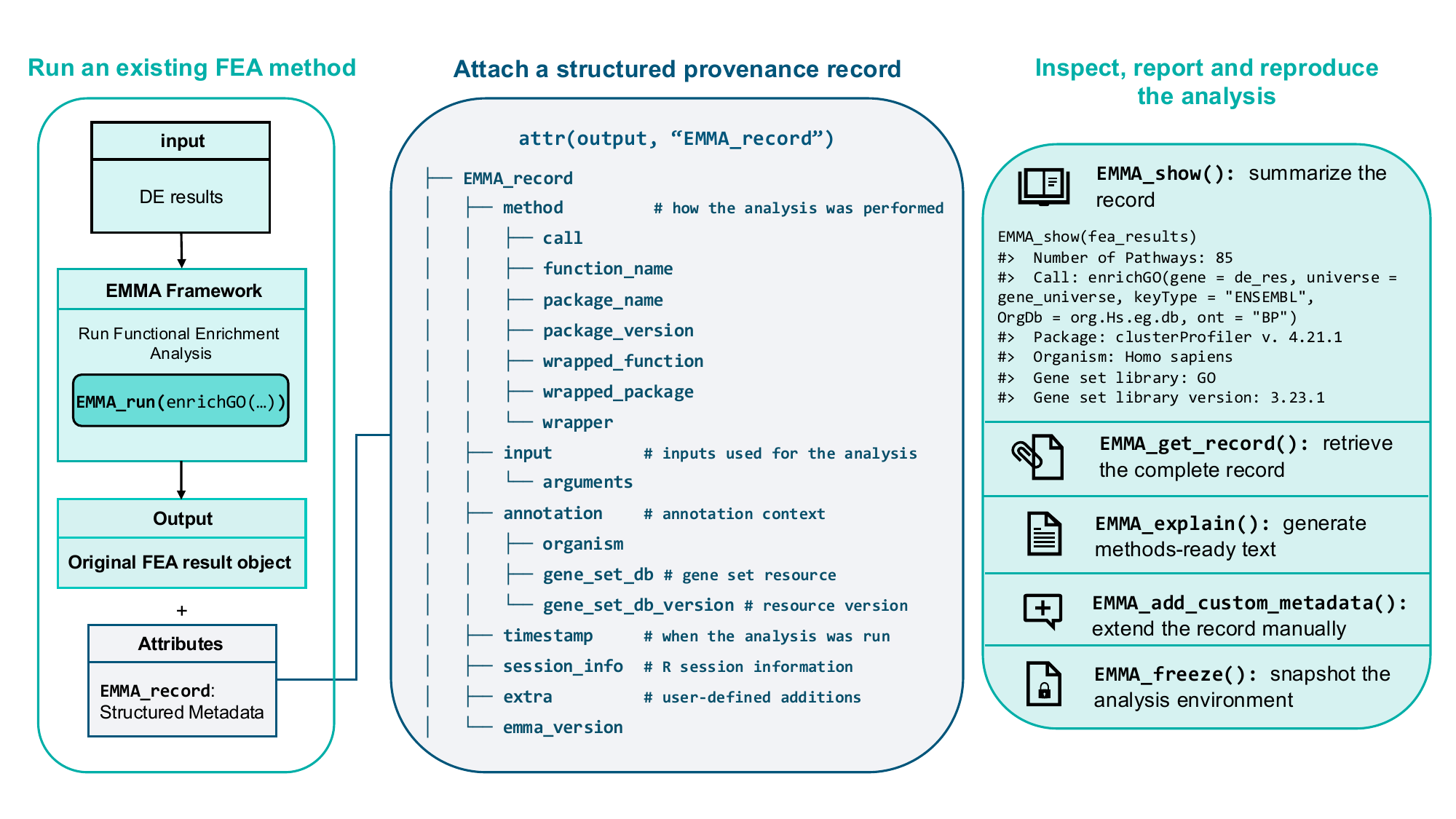}
\caption{EMMA framework for Functional Enrichment Analysis (FEA) and automatic metadata capture. \texttt{EMMA\_run()} evaluates an existing enrichment call (in this example \texttt{enrichGO(...)}) and returns the native result object with an attached structured \texttt{EMMA\_record}. The record captures method, input, and annotation-related information, session, timestamp, and optional user-defined metadata. \texttt{EMMA\_show()} provides a concise summary of the captured provenance. Additional functions allow the user to retrieve, extend, and translate the record into methods-ready text, as well as to create a lockfile providing a detailed snapshot of the analysis environment.
}
\label{fig1}
\end{figure*}

EMMA is designed to integrate seamlessly into existing FEA workflows without requiring any modifications. Rather than replacing existing enrichment tools, EMMA records provenance metadata during analysis execution while preserving existing workflows, and returning the original enrichment results in their native format. The overall EMMA workflow and the structure of the detailed metadata record are illustrated in Fig \ref{fig1}.

\subsection{The core function: \texttt{EMMA\_run()}}

The \texttt{EMMA\_run()} function takes as input an enrichment call, which may invoke functions from packages such as clusterProfiler \citep{Yu2012}, topGO \citep{Alexa2006}, gprofiler2 \citep{Kolberg2020}, goseq \citep{Young2010}, or user-defined custom functions. EMMA then evaluates the call while automatically recording provenance metadata (when available) and storing it as a structured list named \texttt{EMMA\_record}. Rather than introducing a new results container, EMMA leverages R's existing object attribute system to attach \texttt{EMMA\_record} directly to the original enrichment result object, thereby preserving its native format.

The recorded provenance as an \texttt{EMMA\_record} is organized into three main categories: (i) The \emph{methods metadata}, such as the function, the package, software version, and the full call. (ii) The \emph{input metadata}, which corresponds to the parameters passed explicitly to the call. The user can choose whether to store the actual value of the parameters or just the name to avoid inflating the results object size by storing large objects such as organism level annotation databases (Org.db). (iii) The \emph{annotation metadata}, which corresponds to the organism, the gene set database resource used and its version. Recording this information is essential because annotation resources are continuously updated and curated, and changes in their content have been shown to affect enrichment results and their biological interpretation \citep{Wadi2016, Tomczak2018}. Version information is recorded when an identifiable annotation resource is available. For example, for supported Gene Ontology-based analyses, EMMA can record the installed GO.db package version, while methods querying remote resources dynamically may not expose a corresponding version. Therefore, complete provenance capture depends on the version information exposed by the underlying resource. Overall, annotation-specific metadata are inferred when the underlying enrichment function or a supported wrapper can be identified. \\
For custom functions, EMMA captures generic provenance such as the executed call and input metadata, while annotation-specific information can be added manually if it cannot be inferred.

In addition, \texttt{EMMA\_run()} records other metadata such as the R session information and the analysis timestamp. Although some of this information can be documented manually using functions such as \texttt{sessionInfo()} and \texttt{match.call()}, manual provenance capture requires additional technical effort, and is often incomplete. In addition, functions like \texttt{sessionInfo()} do not capture the full analytical context. It mainly describes the software environment, but not necessarily the used parameters. Furthermore, metadata stored separately from the analysis output may become detached from the result to which it applies. EMMA addresses these limitations by automatically capturing provenance at runtime and attaching a structured record directly to the native enrichment result object.

\subsection{Other features}

EMMA also provides utilities to access and export the recorded metadata. Because provenance records may contain extensive metadata, \texttt{EMMA\_show()} provides a concise summary of the result's associated record. The complete provenance record can be retrieved using \texttt{EMMA\_get\_record()}. 

When provenance information cannot be inferred automatically (for example, when executing bespoke analysis scripts), users can extend the record using \texttt{EMMA\_add\_custom\_metadata()} by providing the missing provenance metadata for custom functions or adding supplementary metadata to supported enrichment functions.

To facilitate reporting, EMMA provides \texttt{EMMA\_explain()}, which automatically generates a human-readable description of the enrichment analysis suitable for inclusion in Materials and Methods sections. This is particularly useful in collaborative settings, where sharing the result object together with its attached provenance allows collaborators to obtain a concise summary of how the analysis was performed without having to inspect the underlying metadata or reconstruct the workflow from code or separate documentation. Since the captured record is attached to the FEA results object, it is preserved when the original object is serialized (for example using \texttt{saveRDS()}).

Beyond documenting the analysis itself, EMMA also provides means to support computational reproducibility by defining isolated and portable package environments through \texttt{EMMA\_freeze()}. This function generates a lockfile that can be used with tools such as renv to recreate the analysis environment. We demonstrate this workflow in a supplementary vignette by reproducing Gene Ontology (GO) and Kyoto Encyclopedia of Genes and Genomes (KEGG) enrichment analyses in an isolated Docker container (available at \url{https://github.com/imbeimainz/EMMA_Supplement}).

EMMA is designed to integrate naturally with the Bioconductor ecosystem, to stay consistent with interoperable data structures, reproducible analyses, and comprehensive documentation \citep{Huber2015}. In particular, provenance records can be attached to the original enrichment results object, which then can be stored alongside differential expression analysis results and raw assay data within a container such as a SummarizedExperiment, or more specialized classes such as a DeeDeeExperiment \citep{Abassi2026}, providing a unified representation of the complete analysis workflow.

\section{Conclusion}
EMMA supports the methodological transparency and reproducibility of functional enrichment analyses by automatically capturing provenance metadata without altering existing workflows, while simplifying auditing of existing analyses.

Future versions of EMMA are planned to expand the range of supported enrichment functions and provide more context-aware warnings when key methodological choices, such as the background gene universe or multiple-testing correction procedure, are missing or potentially misspecified. These warnings will be designed to encourage adherence to established best-practice guidelines for FEA. The provenance-aware approach implemented in EMMA could be reapplied in other fields where researchers face a multitude of analysis options, especially in ecosystems where many packages and methodological choices make complete documentation challenging.

\section{Competing interests}
No competing interest is declared.

\section{Author contributions statement}
N.A. Conceptualization, data curation, methodology, software, validation, writing – original draft, writing – review \& editing
A.N. data curation, methodology, software, writing – review \& editing
F.M. Conceptualization, methodology, software, supervision, funding acquisition, project administration, resources, writing – review \& editing.
All authors approved the final version of the manuscript.

\section{Acknowledgments}
The authors thank the Bioconductor community for valuable feedback and suggestions, and also thank Lea Schwarz, Ahmed Hassan, Alicia Schulze, and Alina Jenn for the insightful discussions and constructive suggestions on features and functionality, as well as for testing early versions of the package.

\section{Funding}
This work was supported by the Deutsche Forschungsgemeinschaft (DFG, German Research Foundation) Projektnummer 318346496 - SFB1292/3 TP19N (to NA and FM).

\bibliographystyle{abbrvnat}
\bibliography{references_EMMA}

\end{document}